\documentclass[10pt,letterpaper,twocolumn]{article} 

\usepackage{ol2}
\usepackage[draft]{hyperref}
\usepackage{amsmath}

\begin{document}

\twocolumn[ 

\title{Photonic Bloch oscillations of correlated particles}


\author{Stefano Longhi}

\address{Dipartimento di Fisica, Politecnico di Milano, Piazza L. da Vinci 32, I-20133 Milano, Italy}

\begin{abstract}
A photonic realization of Bloch oscillations (BOs) of two correlated electrons that move on a one-dimensional periodic lattice, based on spatial light transport in a square waveguide array with a defect line, is theoretically proposed.  The 
signature of correlated BOs, such as frequency doubling of the oscillation frequency induced by particle interaction,  can be simply visualized by monitoring the spatial path followed by an optical beam that excites the array near the defect line.
\end{abstract}



 ] 

\noindent 
Bloch oscillations (BOs), i.e.,  the oscillatory motion of
electrons in a periodic periodical induced by a dc field, are
one of the most striking predictions of the semiclassical
theory of electronic transport. BOs manifest the
wave properties of the electrons, and therefore appear in
other systems of waves in tilted periodic potentials. BOs
have been observed for electrons in semiconductor
superlattices, matter waves in optical lattices,
and acoustic or light waves in periodic media. In optics,  
analogues of BOs occur in different structures, including 
dielectric waveguide arrays \cite{a1,a2,a3,a4,a5,a6}, optical superlattices \cite{s1,s2} and metal-dielectric structures \cite{p1,p2,p3}.
 Quantum 
signatures of nonclassical light undergoing BOs have been investigated as well \cite{Longhi08,Rai09,Lahini10}.
One of the main limitations of  photonic BOs so far realized is    
to mimic the motion of {\em  single} particles solely. For many particles, the onset of BOs is greatly affected by 
particle interactions \cite{i1,i1bis,i3,i4,i5}, and interesting novel phenomena are predicted for BOs of {\em few} interacting particles \cite{i1bis,i3,i5}, such as
the frequency doubling of  BOs of two correlated  electrons \cite{i1bis,i3}. Experiments aimed to observe BOs of {\em few} correlated electrons or bosons are rather difficult to be performed in semiconductor superlattices or cold atoms, where 
the many particle regime is generally of easier access. 
In this Letter a photonic realization of  BOs for two correlated 
electrons is proposed, which is based on light transport in a two-dimensional square waveguide array 
with a defect line.  \par
The motion of two interacting electrons moving  on a one-dimensional tight-binding lattice subjected to an external dc force $F$ is described by the
Hubbard Hamiltonian (see, for instance, \cite{i1bis,i3})
\begin{eqnarray}
\hat{H} &=& -\kappa \sum_{n,s=\uparrow,\downarrow}  \left(\hat{a}^{\dag}_{n+1,s}\hat{a}_{n,s}+ \hat{a}^{\dag}_{n,s}\hat{a}_{n+1,s} \right)  \\
& + &   \sum_{n,s=\uparrow,\downarrow} 
Fa\hat{n}\hat{a}^{\dag}_{n,s}\hat{a}_{n,s}- U \sum_n \hat{a}^{\dag}_{n,\uparrow} \hat{a}_{n,\uparrow}\hat{a}^{\dag}_{n,\downarrow} \hat{a}_{n,\downarrow} \nonumber
\end{eqnarray}
 where $\hat{a}_{n,s}$ and $\hat{a}^{\dag}_{n,s}$ are the annihilation and creation operators for the electron at site $n$ with spin $s=\uparrow,\downarrow$, $\hat{n}$ is the position operator, $\kappa$ is the hopping amplitude, $a$ is the lattice period, and $U$ is the on-site electron-electron interaction strength. In order to allow for double occupancy of the on-site orbital, the two electrons are assumed to have opposite spins (singlet state). A photonic realization of the Hubbard Hamiltonian (1) for two electrons can be readily obtained after expanding the state vector of the system $|\psi(t) \rangle$ as a superposition of Wannier states, i.e. $|\psi(t) \rangle=\sum_{n,m}c_{n,m}(t) |ns_1,ms_2 \rangle$, where the ket $|ns_1,ms_2 \rangle$  represents a state with one electron with spin $s_1$ at site $n$ and the other electron with spin $s_2$ at site $m$. In the Wannier representation, the time evolution of the quantum state $|\psi(t) \rangle$, governed by the Schr\"odinger equation with $\hbar=1$, i.e. $i \partial_t |\psi(t) \rangle=\hat{H}|\psi(t) $, reads explicitly
 \begin{eqnarray}
 i \frac{dc_{n,m}}{dt} & = & - \kappa ( c_{n+1,m}+c_{n-1,m}+c_{n,m+1}+c_{n,m-1}) + \nonumber \\
 & + & [Fa(n+m)-U \delta_{n,m}]c_{n,m}.
 \end{eqnarray}
\begin{figure}[htb]
\centerline{\includegraphics[width=8cm]{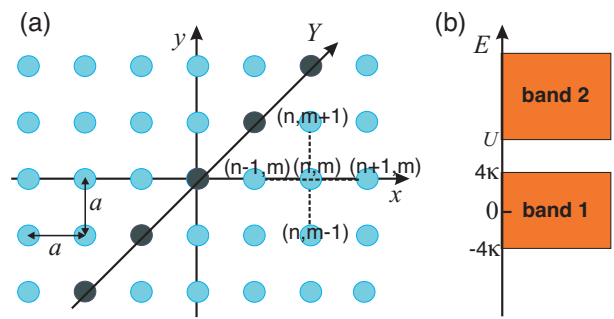}} \caption{
(Color online) (a) Schematic of a square waveguide array with a line defect that realizes the Hubbard Hamiltonian (1) for two electrons with opposite spins. (b) Band structure of the tight-binding lattice for a nonvanishing interaction and in the absence of the dc force.}
\end{figure}
In their present form, Eqs.(2) can be viewed as the coupled-mode equations describing light transport in a two-dimensional square array of waveguides with a superimposed transverse index gradient and with a defect line (interface) at the diagonal lattice sites $n=m$, in which the hopping amplitude $\kappa$ defines  the coupling constant between adjacent waveguides, the on-site electron interaction $U$ defines the propagation constant shift of the guided modes for the waveguides along the interface  $n=m$, and the temporal evolution of the Wannier amplitudes $c_{n,m}$ is mapped into the spatial evolution of the modal amplitudes of light waves, along the array axis, trapped in the various waveguides [see Fig.1(a)]. Such a structure could be realized, for instance, by femtosecond laser writing in fused silica, in which the defect line is readily obtained by varying the writing speed of the laser beam (see, for instance, \cite{a6,R2}). The transverse index gradient is achieved by circularly-curving the waveguides in the $(Y,z)$ plane, where the $Y$ direction defines the defect line [see Fig.1(a)] . In the waveguide reference frame, propagation of light waves at wavelength $\lambda$ is described by the optical Schr\"odinger equation for the electric field envelope $\phi(x,y,s)$ \cite{R3}
\begin{equation}
i \hbar \frac{\partial \phi}{\partial s}=-\frac{\hbar^2}{2n_s}\nabla^2_{x,y} \phi+V(x,y) \phi+\frac{n_s (x+y)}{\sqrt 2 R}  \phi
\end{equation}
where $\hbar= \lambda/(2 \pi)$ is the reduced wavelength of light, $s$ is the curvilinear coordinate along the axis of a reference waveguide, $n_s$ is the substrate refractive index, $R$ is the radius of curvature of waveguides, and $V(x,y) \simeq n_s-n(x,y)$ is the optical potential which is determined by the refractive index change $n_s-n(x,y)$ inscribed in the substrate. Indicating by $h(x,y)$ the normalized index profile of each waveguide core and by $\Delta n$, $\Delta n_1$ the index changes realized in the lattice   and along the defect line, one can write $V(x,y)=-\sum_{n,m} [\Delta n+(\Delta n_1-\Delta n) \delta_{n,m}]h(x-na,y-ma)$, where  $a$ is the lattice period.  Coupled-mode equations (2) are obtained from Eq.(3) in the tight-binding approximation and neglecting cross coupling, provided that   time $t$ is replaced by the spatial curvilinear  distance $s$ and a forcing  $F=2 \pi n_s/( \sqrt 2 R \lambda)$ is assumed. \par
To understand the appearance of frequency doubling in the BOs of two interacting electrons, it is worth considering the energy spectrum of the two-electron Hamiltonian in the absence of the external force, i.e. for $F=0$ \cite{i1bis,i3}.  For $U=0$, there exists a single tight-binding band of energy $-4 \kappa<E<4 \kappa$, which is precisely the energy band of a {\em single} electron on a two-dimensional square lattice. The corresponding eigenfunctions are Bloch states which are fully delocalized in the crystal. For a non-vanishing interaction $U$,  the energy spectrum is composed by two bands \cite{i1bis}. The first one, which covers the range $-4 \kappa<E<4 \kappa$ [band 1 in Fig.1(b)], corresponds to delocalized Bloch states which are scattered off by the defect line $n=m$. The second band [band 2 in Fig.1(b)],  which covers the range $ U <E< \sqrt{U^2+16 \kappa^2} $, is 
related to the appearance of {\em defect modes} which are localized at the $n=m$ interface [i.e. around the $Y$ axis of Fig.1(a)]. These two bands starts to separate when the interaction $U$ increases above $4 \kappa$ [see Fig.1(b)]. The initial distribution $c_{n,m}(t=0)$ of the two electrons in the lattice is assumed to be Gaussian-shaped, i.e   $c_{n,m}(t=0)= Z \exp[-(n-n_0)^2/w^2-(m-m_0)^2/w^2]$, were $Z$ is the normalization factor, $n_0$ and $m_0$ are the mean positions of the two electrons in the lattice, and $w$ measures the localization length (in units of the lattice period $a$) of the electronic wave function. Such an initial condition is readily realized in the optical system by initial excitation of the array, at the $s=0$ input plane,  with a Gaussian beam of spot size $wa$ with normal incidence at the lattice sites $(n_0,m_0)$, i.e. 
$\phi(x,y,0)=\exp[-(x-n_0a)^2/(wa)^2-(y-m_0a)^2/(wa)^2]$. 
Note that, if $n_0=m_0$, the wave packet describes two electrons that initially occupy  the same site and thus strongly interact. Conversely, if $|n_0-m_0|$ is much larger than $w$, the initial wave packet describes two electrons that are initially separated each other. When the external force $F$ is applied,  in the latter case the two particles basically undergo independent BOs with a characteristic period given by $T_B=2 \pi/(Fa)$, provided that the amplitude of BOs is smaller than the electronic separation.  In the optical lattice realization of Fig.1(a), such a result can be simply explained by observing that in this case the two-dimensional wave packet motion basically remains confined in a homogeneous region of the lattice and does not touches the defect line $n=m$, thus realizing a two-dimensional BOs motion. Conversely, if the two electrons initially occupy the same site, i.e. for $n_0=m_0$, BOs develop a
frequency doubled component which is more pronounced at
intermediate couplings. As discussed in Ref. \cite{i3}, such frequency doubling is associated
with the excitation of bounded states of two electrons
in a singlet configuration (i.e. the localized modes at the defect line $m=n$).  The relative contributions 
of fundamental and frequency double components of the BOs basically depends on the relative
excitation of bounded and unbounded
states from the initial wave packet. In particular, the frequency-doubled component is more pronounced for $U \sim 4 \kappa$, i.e. at the interaction strength at which the two lattice bands starts to separate each other, whereas it vanishes in the low and strong coupling limits \cite{i3}.  The onset of independent BOs for spatially-separated electrons, and of correlated BOs with frequency doubling of the oscillation frequency for closely-spaced interacting electrons, is shown in Figs.2 and 3, respectively.  The figures show results obtained by numerical simulations of the paraxial wave equation (3)   in an array of length $5$ cm for parameter values $\lambda=980$ nm, $n_s=1.522$, $\Delta n=0.01$, $\Delta n_1=0.01035$, $a=8.6 \; \mu$m, $R= (30/\sqrt 2)$ cm and for a Gaussian profile $h(x,y)$ of the waveguide core of radius $3 \; \mu$m.  The values of coupling constant $\kappa$, propagation constant shift $U$ and gradient parameter $Fa$ entering in the coupled-mode equations (2) are estimated to be $\kappa=3.977 \; {\rm cm}^{-1}$, $U \simeq 4 \kappa$, and $Fa=2.7973 \; {\rm cm}^{-1}$,  respectively. The spatial period of BOs is $T_B=2 \pi/(Fa) \simeq 2.25$ cm. Parameter values of the input Gaussian beam are $n_0=10$, $m_0=0$,$w=1.3$ in Fig.2, and $n_0=0$, $m_0=0$,$w=1.3$ in Fig.3. The dotted curves in Figs.2(a) and 3(a) show the numerically-computed evolution of the beam centroid $\langle n (s) \rangle$, defined by the relation  
\begin{equation}
\langle n (s) \rangle=\frac{ \int dx dy (x/a-n_0)|\phi(x,y,s)|^2 }{ \int dx dy |\phi(x,y,s)|^2},
\end{equation}
 clearly showing the doubling of the oscillation frequency when the lattice is initially excited at the defect line [compare Figs.2(a) and 3(a)].  The curves are well approximated by the corresponding ones for the wave packet centroid $\langle n (t) \rangle=\sum_{n,m}(n-n_0)|c_{n,m}(t)|^2$ obtained by numerical simulations of the coupled mode equations (2), which are depicted by the solid curves in the figures. Snapshots of $|\phi(x,y,s)|$ at a few propagation distances $s$ are also shown in Figs. 2(b) and 3(b). Note that, while for two electrons initially far apart each other the wave packet undergoes two-dimensional BOs   
in a homogeneous region of the array, without touching the defect line $Y$  [see Fig.2(b)], for the two interacting electrons the wave packet remains strongly localized near the defect line, exciting the defect modes at the interface  [Fig.3(b)]. \par
In conclusion, a photonic realization of BOs for two correlated electrons has been proposed, which is based on light transport in a square waveguide array with a defect line. Such an optical setting should provide an experimentally accessible laboratory tool for the observation of interaction-induced frequency doubling of BOs.
\begin{figure}[htb]
\centerline{\includegraphics[width=8cm]{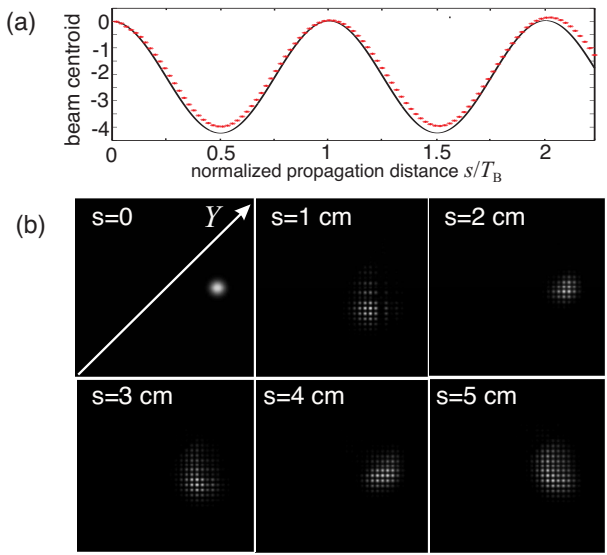}} \caption{
(Color online) BOs for two initially-separated (non-interacting) electrons. (a) Evolution of the beam centroid versus the normalized propagation length $s/T_B$, and (b) snapshots of $|\phi(x,y,s)|$ at a few propagation distances $s$.}
\end{figure}

\begin{figure}[htb]
\centerline{\includegraphics[width=8cm]{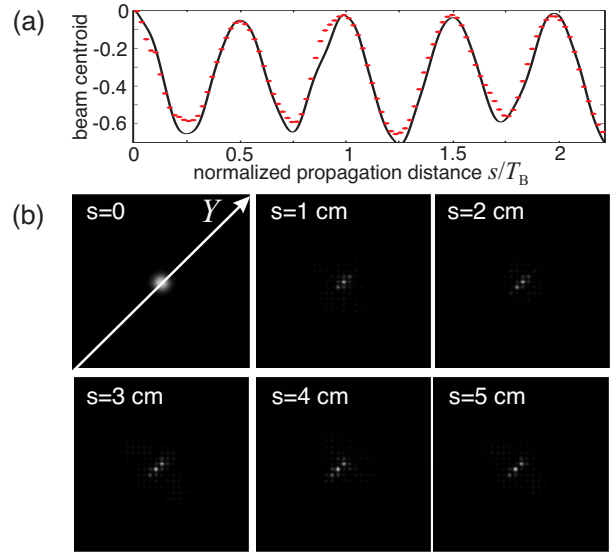}} \caption{
(Color online) Same as Fig.2, but for interacting electrons.}
\end{figure}

\par
Work supported by the Italian MIUR (Grant No. PRIN-2008-YCAAK).

\newpage

\footnotesize {\bf References with full titles}\\
\\

1. T. Pertsch, P. Dannberg, W. Elflein, A. Br\"{a}uer, and F.
Lederer, "Optical Bloch Oscillations in Temperature Tuned Waveguide
Arrays", Phys. Rev. Lett. {\bf 83}, 4752 (1999).\\

2. R. Morandotti, U. Peschel, J. Aitchinson, H. Eisenberg, and Y.
Silberberg, "Experimental observation of linear and nonlinear
optical bloch oscillations", Phys. Rev. Lett. {\bf 83}, 4756
(1999).\\

3. H. Trompeter, T. Pertsch, F. Lederer, D. Michaelis, U. Streppel,
A. Br\"{a}uer, and U. Peschel, "Visual Observation of Zener
Tunneling," Phys. Rev. Lett. {\bf 96}, 023901 (2006).\\

4. H. Trompeter, W. Krolikowski, D.N. Neshev, A.S. Desyatnikov, A.A. Sukhorukov, Y.S. Kivshar, T. Pertsch, U. Peschel, and F. Lederer,
"Bloch Oscillations and Zener Tunneling in Two-Dimensional Photonic Lattices", Phys. Rev. Lett. {\bf 96}, 053903  (2006).\\

5. N. Chiodo, G. Della Valle, R. Osellame, S. Longhi, G. Cerullo, R. Ramponi, P. Laporta, and U. Morgner,
 "Imaging of Bloch oscillations in erbium-doped curved waveguide arrays", Opt. Lett. {\bf 31}, 1651 (2006).\\

6. F. Dreisow, A. Szameit, M. Heinrich, T. Pertsch, S. Nolte, A.
T\"{u}nnermann, and S. Longhi, "Bloch-Zener Oscillations in Binary
Superlattices", Phys. Rev. Lett. {\bf 102}, 076802 (2009).\\

7. R. Sapienza, P. Costantino, D. Wiersma, M. Ghulinyan, C.J. Oton, and L. Pavesi, "Optical Analogue of Electronic Bloch Oscillations", Phys. Rev. Lett. {\bf 91}, 263902 (2003). \\

8. V. Agarwal, J. A. del R'o, G. Malpuech, M. Zamfirescu, A. Kavokin, D. Coquillat, D. Scalbert, M. Vladimirova, and B. Gil, "Photon Bloch Oscillations in Porous Silicon Optical Superlattices", Phys. Rev. Lett. {\bf 92}, 097401 (2004).\\

9. W. Lin, Z. Xiang, G. P. Wang, and C. T. Chan, "Spatial Bloch oscillations of plasmons in nanoscale metal waveguide arrays ", Appl. Phys. Lett. {\bf 91}, 243113 (2007).\\

10. A.R. Davoyan, I.V. Shadrivov, A.A. Sukhorukov, and Y.S. Kivshar, "Plasmonic Bloch oscillations in chirped metal-dielectric structures", Appl. Phys. Lett. {\bf 94}, 161105 (2009).\\

11. R.-C. Shiu, Y.-C. Lan, and C.-M. Chen, "Plasmonic Bloch oscillations in cylindrical metalÐdielectric waveguide arrays ", Opt. Lett. {\bf 35}, 4012 (2010).\\

12. S. Longhi, "Optical Bloch Oscillations and Zener Tunneling with Nonclassical Light", Phys. Rev. Lett. {\bf 101},  193902 (2008).\\

13. A. Rai and G.S. Agarwal, "Possibility of coherent phenomena such as Bloch oscillations with single photons via W states", Phys. Rev. A {\bf 79}, 053849 (2009). \\

14. Y. Bromberg, Y. Lahini, and Y Silberberg, 
"Bloch Oscillations of Path-Entangled Photons",
Phys. Rev. Lett. {\bf 105}, 263604 (2010).\\

15. A. Buchleitner and A.R. Kolovsky, "Interaction-induced decoherence of atomic Bloch oscillations", Phys. Rev. Lett. {\bf 91}, 253002 (2003).\\

16. F. Claro, J. F. Weisz, and S. Curilef, "Interaction-induced oscillations in correlated electron transport", Phys. Rev. B {\bf 67}, 193101 (2003).\\

17. W. S. Dias, E. M. Nascimento, M. L. Lyra, and F. A. B. F. de Moura, "Frequency doubling of Bloch oscillations for interacting electrons in a static electric field", Phys. Rev. B {\bf 76}, 155124 (2007). \\

18. J. K. Freericks, "Quenching Bloch oscillations in a strongly correlated material: Nonequilibrium dynamical mean-field theory
", Phys. Rev. B {\bf 77}, 075109 (2008).\\

19. R. Khomeriki, D.O. Krimer, M. Haque, and S. Flach, "Interaction-induced fractional Bloch and tunneling oscillations",
Phys. Rev. A {\bf 81}, 065601 (2010). \\

20. A. Szameit and S. Nolte, "Discrete optics in
femtosecond-laser-written photonic structures," J. Phys. B {\bf 43},
163001 (2010).\\

21. S. Longhi, "Quantum-optical analogies using photonic structures",
Laser and Photon. Rev. {\bf 3}, 243 (2009).


\begin{thebibliography}{99}



\bibitem{a1}
T. Pertsch, P. Dannberg, W. Elflein, A. Br\"{a}uer, and F.
Lederer,  Phys. Rev. Lett. {\bf 83}, 4752 (1999).

\bibitem{a2}
R. Morandotti, U. Peschel, J. Aitchinson, H. Eisenberg, and Y.
Silberberg, Phys. Rev. Lett. {\bf 83}, 4756
(1999).

\bibitem{a3}
H. Trompeter, T. Pertsch, F. Lederer, D. Michaelis, U. Streppel,
A. Br\"{a}uer, and U. Peschel, Phys. Rev. Lett. {\bf 96}, 023901 (2006).

\bibitem{a4}
H. Trompeter, W. Krolikowski, D.N. Neshev, A.S. Desyatnikov, A.A. Sukhorukov, Y.S. Kivshar, T. Pertsch, U. Peschel, and F. Lederer,
Phys. Rev. Lett. {\bf 96}, 053903  (2006).

\bibitem{a5}
N. Chiodo, G. Della Valle, R. Osellame, S. Longhi, G. Cerullo, R. Ramponi, P. Laporta, and U. Morgner,
 Opt. Lett. {\bf 31}, 1651 (2006).

\bibitem{a6}
 F. Dreisow, A. Szameit, M. Heinrich, T. Pertsch, S. Nolte, A.
T\"{u}nnermann, and S. Longhi, Phys. Rev. Lett. {\bf 102}, 076802 (2009).

\bibitem{s1}
R. Sapienza, P. Costantino, D. Wiersma, M. Ghulinyan, C.J. Oton, and L. Pavesi,  Phys. Rev. Lett. {\bf 91}, 263902 (2003). 

\bibitem{s2}
V. Agarwal, J. A. del R'o, G. Malpuech, M. Zamfirescu, A. Kavokin, D. Coquillat, D. Scalbert, M. Vladimirova, and B. Gil, Phys. Rev. Lett. {\bf 92}, 097401 (2004).

\bibitem{p1}
W. Lin, Z. Xiang, G. P. Wang, and C. T. Chan, Appl. Phys. Lett. {\bf 91}, 243113 (2007).

\bibitem{p2}
A.R. Davoyan, I.V. Shadrivov, A.A. Sukhorukov, and Y.S. Kivshar, Appl. Phys. Lett. {\bf 94}, 161105 (2009).

\bibitem{p3}
R.-C. Shiu, Y.-C. Lan, and C.-M. Chen, Opt. Lett. {\bf 35}, 4012 (2010). 

\bibitem{Longhi08}
S. Longhi, Phys. Rev. Lett. {\bf 101},  193902 (2008).

\bibitem{Rai09}
A. Rai and G.S. Agarwal, Phys. Rev. A {\bf 79}, 053849 (2009). 

\bibitem{Lahini10}
Y. Bromberg, Y. Lahini, and Y Silberberg, 
Phys. Rev. Lett. {\bf 105}, 263604 (2010).

\bibitem{i1}
A. Buchleitner and A.R. Kolovsky, Phys. Rev. Lett. {\bf 91}, 253002 (2003).

\bibitem{i1bis}
F. Claro, J. F. Weisz, and S. Curilef, Phys. Rev. B {\bf 67}, 193101 (2003).

\bibitem{i3}
W. S. Dias, E. M. Nascimento, M. L. Lyra, and F. A. B. F. de Moura, Phys. Rev. B {\bf 76}, 155124 (2007).
 
\bibitem{i4}
J. K. Freericks, Phys. Rev. B {\bf 77}, 075109 (2008).

\bibitem{i5}
R. Khomeriki, D.O. Krimer, M. Haque, and S. Flach, 
Phys. Rev. A {\bf 81}, 065601 (2010).

\bibitem{R2}
A. Szameit and S. Nolte,  J. Phys. B {\bf 43},
163001 (2010).

\bibitem{R3}
S. Longhi, 
Laser and Photon. Rev. {\bf 3}, 243 (2009).





%
%
%
%
%
%
%
%
%
%
%
%
%
%
%
%
%
%
%
%
%
\end{thebibliography}
\end{document}